\documentclass[reprint,superscriptaddress, amsmath,amssymb, aps, pra, longbibliography]{revtex4-1}

\usepackage{silence}
\usepackage{soul}
\usepackage{amsmath}   
\usepackage{amssymb}
\usepackage{mathtools}
\usepackage{graphicx}
\usepackage{dcolumn}
\usepackage{bm}
\usepackage{amsfonts}
\usepackage{subfigure}
\usepackage{array}
\usepackage{float}
\usepackage{color}
\usepackage{multirow}
\usepackage{makecell}
\usepackage{amssymb}
\usepackage[colorlinks=true,linkcolor=blue]{hyperref}
\hypersetup{allcolors=blue}
\usepackage[normalem]{ulem}
\usepackage{xcolor}

\usepackage{mathtools}
\DeclarePairedDelimiterX\braket[2]{\langle}{\rangle}{#1 \delimsize\vert #2}

\begin{document}
\title{SI-Traceable Calibration and Performance Benchmarking of a Terahertz Photomixer Transmitter--Receiver System Using a Rydberg Atomic Sensor}

\date{\today }

\author{Alisher Duspayev}
    \email{ADuspayev@RydbergTech.com}
    \affiliation{Rydberg Technologies Inc., Ann Arbor, Michigan 48103, USA}
\author{Kaitlin R. Moore}
    \affiliation{Rydberg Technologies Inc., Ann Arbor, Michigan 48103, USA}
\author{Georg Raithel}
    \affiliation{Rydberg Technologies Inc., Ann Arbor, Michigan 48103, USA}
\author{David A. Anderson}
    \affiliation{Rydberg Technologies Inc., Ann Arbor, Michigan 48103, USA}    

\begin{abstract}
Accurate calibration of electromagnetic field strength in the terahertz (THz) frequency regime remains challenging due to the limited availability of SI-traceable field sensors. Here we demonstrate SI-traceable calibration and performance benchmarking of a commercial photomixer-based THz transmitter--receiver system operating near 204~GHz using Rydberg electric-field sensing in a thermal atomic vapor. The THz electric field is extracted from the Autler-Townes (AT) effect of the cesium $17D_{5/2}\rightarrow18P_{3/2}$ Rydberg transition. We measure the strength of AT-split lines as a function of THz detuning from resonance to find the on-resonant Rabi frequency, which, together with atomic transition dipole moments and fundamental constants, yields the THz electric field. The atomically measured field calibrates the photomixer transmitter field and power, while simultaneous measurements with a commercial InGaAs photomixer receiver calibrate the receiver's current-to-field responsivity and convert its current-noise floor into an absolute noise-equivalent THz electric-field sensitivity. Our study demonstrates that Rydberg atomic sensors provide a practical method for SI-traceable calibration and benchmarking of THz transmitter and receiver systems, and for the establishment of a quantitative link between state-of-the-art and absolute atom-based THz sensors. 
\end{abstract}

\maketitle

\section{Introduction}
\label{sec:intro}
Terahertz (THz) and millimeter-wave technologies are critical platforms for sensing, imaging, spectroscopy, communications, manufacturing, and metrology~\cite{Davies_2002, Tonouchi2007, Lewis_2014}, underpinning applications in materials science~\cite{Shi_2023}, food and agriculture~\cite{THzfood2019, Gente2015}, medicine~\cite{siegel2004}, precision metrology~\cite{Shang2024, Baxter2011}, and high-frequency communications~\cite{Thomas2025}, among others~\cite{kemp2003, Pickwell_2006, PAWAR2013157}. Accurate and precise measurement of electromagnetic fields in the THz and sub-THz frequency ranges is of fundamental importance for applications of such fields in research and technology. However, calibration of THz transmitters and detectors remains challenging due to the lack of widely available traceable field standards above 100~GHz~\cite{Lewis_2019}.

Photomixer-based THz systems~\cite{preu2011} are widely used as coherent transmitters and receivers in THz applications. These devices generate or detect THz radiation through optical heterodyning of two near-infrared lasers in a photoconductive antenna. The resulting photocurrent oscillates at the optical beat frequency, producing relatively narrow-band radiation within the millimeter-wave or THz frequency ranges. In reception, an incident THz electric field modulates the photoconductive current generated by the optical beat signal. The resulting photocurrent modulation is typically detected using lock-in amplification. The responsivity of photomixer receivers depends on device-specific parameters including antenna impedance, optical pump power, focusing and alignment of the incident THz field, and bias conditions. As a result, the calibration of THz electric field amplitude using photomixer detectors often resorts to indirect power estimates or vendor specifications~\cite{cai1998, Lehman_2012, Yardimci2018, Aji2023, Lin_2020}.

Rydberg atomic sensors~\cite{Anderson2020, fancher2021, Schlossberger2024} provide an alternative approach to electromagnetic field measurement that is inherently traceable to fundamental and atomic constants~\cite{Sedlacek2012, Holloway2014, Holloway2017, AndersonRFMS2021}. Recently, these systems have been utilized for THz imaging~\cite{AndersonGSMM.2018, downes2020, Downes_2023, Li_2025}, electrometry~\cite{Chen:22, Krokosz:25, wang2026} and heterodyne atomic reception~\cite{Legaie.2024, She_2024, LIN2025970, Xing:26}. When a resonant and sufficiently intense radio-frequency (RF) or THz field couples two Rydberg states, the interaction produces an Autler--Townes (AT) splitting equal to the RF Rabi frequency, $\Omega = E_{THz} d / h $ (in Hz), where both $\Omega$ and $E_{THz}$ are taken to be positive and real without loss of generality. Because the electric dipole transition matrix element (EDTM), $d$, between Rydberg states can be calculated with high precision~\cite{Safronova2004, Reinhard2007, SIBALIC2017319, Tran2023}, the electric-field amplitude, $E_{THz}$, can be extracted directly from the spectroscopically measured AT splitting~\cite{Holloway2014}. This self-calibrated method is especially attractive in the THz regime, where transfer-standard calibration approaches are still limited. 

In this work, we demonstrate concurrent measurement of an over-the-air THz field near 204~GHz, generated by a commercial photomixer source, using both a Rydberg atomic sensor and a commercial photomixer receiver. The atomic Cs vapor-cell sensor provides an absolute, SI-calibrated measurement of the THz electric field as a function of bias voltage of the photomixer emitter through the AT response of the $17D_{5/2}\rightarrow18P_{3/2}$ Rydberg transition, which is resonant near 204.52~GHz. The same THz field is concurrently measured using the photomixer receiver. The measurements are performed as a function of the photomixer emitter's dc bias-voltage offset, $V_0$, and its bias-voltage modulation amplitude, $V_{mod}$. The bias-voltage modulation is critical for the photomixer receiver's lock-in detection scheme. Combining the measurements allows for an absolute calibration of the THz photomixer receiver's electric-field responsivity against an atomic-standard method. Thus, our work creates a calibration link between absolute, atom-based, SI-traceable THz field measurement and frequently-used THz photomixer receivers.

\section{Experimental Setup}
\label{sec:setup}

\begin{figure}[t!]
\centering
\includegraphics[width=1\linewidth]{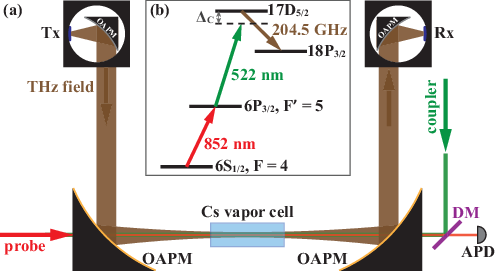}
\caption{(a) Sketch of the experimental setup (not to scale). Tx: transmitter; Rx: receiver; OAPM: off-axis parabolic mirror; DM: dichoric mirror; APD: avalanche photodetector. (b) Single-atom energy-level diagram (not to scale) outlining utilized Cs states.}
\label{fig:setup}
\end{figure}

The experimental setup, sketched in Fig.~\ref{fig:setup}~(a), consists of a commercial photomixer-based THz transmitter and receiver [Tx and Rx in Fig.~\ref{fig:setup}~(a)] system (Toptica Terascan~\cite{terascanpaper2022, TopticaWP}) and a vapor-cell-based spectroscopic Rydberg electrometry system. The Tx produces THz radiation by optical mixing of two distributed-feedback (DFB) lasers [not shown in Fig.~\ref{fig:setup}~(a)]. The frequencies of the DFB lasers are stabilized and tuned by locking them to a high-resolution wavemeter, allowing us to tune their frequency difference (and, hence, the THz field frequency) with $\lesssim$~2~MHz uncertainty. We tune the THz frequency from about 204.4~GHz to 204.65~GHz, which covers the atomic THz resonance that is used for field calibration in the present work. The THz field is emitted into a near-Gaussian beam from the photoconductive antenna in the Tx unit. The THz beam is collimated and focused through a Cs vapor cell using metal-coated off-axis parabolic mirrors [OAPM in Fig.~\ref{fig:setup}~(a)]. The Gaussian beam waist parameter at the THz focal spot is measured to be $w_{THz}=$~5.6(5)~mm. The Tx source power is varied by changing the bias voltage, $V_{bias}$, which is applied to the InGaAs photomixer diode. A matched receiver [Rx in Fig.~\ref{fig:setup}~(a)] detects the THz photocurrent using lock-in detection. Details and the corresponding implications for the calibration procedure are elaborated on in Sec.~\ref{sec:cali}.

The atomic Rydberg electrometry system utilizes counter-propagating probe (852~nm) and coupler (522~nm) laser beams, which drive the Cs $|6S_{1/2},F=4\rangle\rightarrow|6P_{3/2}, F'=5\rangle$ and  $|6P_{3/2},F'=5\rangle\rightarrow|17D_{5/2}\rangle$ transitions, respectively. The linearly polarized THz field co-propagates with the probe beam and is confocal with the optical beams. The THz beam drives the $|17D_{5/2}\rangle\rightarrow|18P_{3/2}\rangle$ Rydberg transition, which has a resonant frequency of $\nu_{THz, 0} = 204.518(2)$~GHz (as measured below) and an EDTM of $\bar{d} = 139 e a_0$ (averaged over the THz-coupled $|m_j| = 1/2$ and $|m_j| = 3/2$ states). The relevant energy-level diagram is depicted in Fig.~\ref{fig:setup}~(b). The probe is frequency-locked to the $|6S_{1/2}, F=4\rangle\rightarrow|6P_{3/2}, F'=5\rangle$ transition using saturation absorption spectroscopy in an auxiliary Cs vapor cell [not shown in Fig.~\ref{fig:setup}~(a)]. The probe beam power transmitted through the 7.5-cm-long Pyrex vapor cell shown in Fig.~\ref{fig:setup}~(a) (main vapor cell), $P_P$, is measured using an avalanche photodetector [APD in Fig.~\ref{fig:setup}~(a)]. The coupler frequency detuning from the THz-field-free $|6P_{3/2},F'=5\rangle\rightarrow|17D_{5/2}\rangle$ transition, $\Delta_C$, is scanned over a range of 300~MHz, centered at the transition. 
The recorded spectra $P_P(\Delta_C)$ exhibit Rydberg-EIT lines. We measure the THz-induced AT-splitting behavior of the EIT lines. The reference point $\Delta_C = 0$ is determined by simultaneous acquisition of THz-field-free EIT spectra in another auxiliary vapor cell [not shown in Fig.~\ref{fig:setup}~(a)]. The polarizations of the optical and THz beams are linear and parallel to each other. The probe and coupler-laser powers measured before the main cell are approximately 46~$\mu$W and 21~mW, respectively, and their Gaussian beam waist radii are approximately 0.6~mm and 0.7~mm, corresponding to respective Rabi frequencies of approximately 4~MHz and 5~MHz.

\section{Terahertz System Calibration}
\label{sec:cali}

In the utilized THz system, both photomixer Tx and Rx are fed by identical DFB lasers via fiber couplers and splitters (see Sec.~\ref{fig:setup} for details on the DFB-laser frequency control). When the THz field emitted by the Tx is incident on the Rx, the latter produces a photocurrent with a magnitude that is proportional to the THz field strength, $E_{THz}$ (notably, not the THz power, $P_{THz}$)~\cite{AD052026}. To afford low signal level detection, the Tx bias voltage $V_{bias}$ can be modulated with a sine wave for lock-in detection of the Rx signal. In this default operation mode, the bias voltage applied to the Tx photomixer is

\begin{equation}
V_{bias} = V_0 + V_{mod}\cos{(\omega_{mod} t+\phi)}\quad,
\label{eq:Vbias}
\end{equation}

\noindent where $V_0$ is a fixed, user-selectable bias voltage (typically, between -1 and 0~V), $V_{mod}$ is the modulation amplitude, and $\omega_{mod}$ and $\phi$ are the angular frequency and phase offset of the modulating sine wave. The two latter parameters were set upon commissioning of the system and were not varied in the present work. 

\subsection{Continuous mode}
\label{subsec:calidc}

\subsubsection{Method}
\label{subsubsec:calidcmethod}

\begin{figure}[t]
\centering
\includegraphics[width=0.98\linewidth]{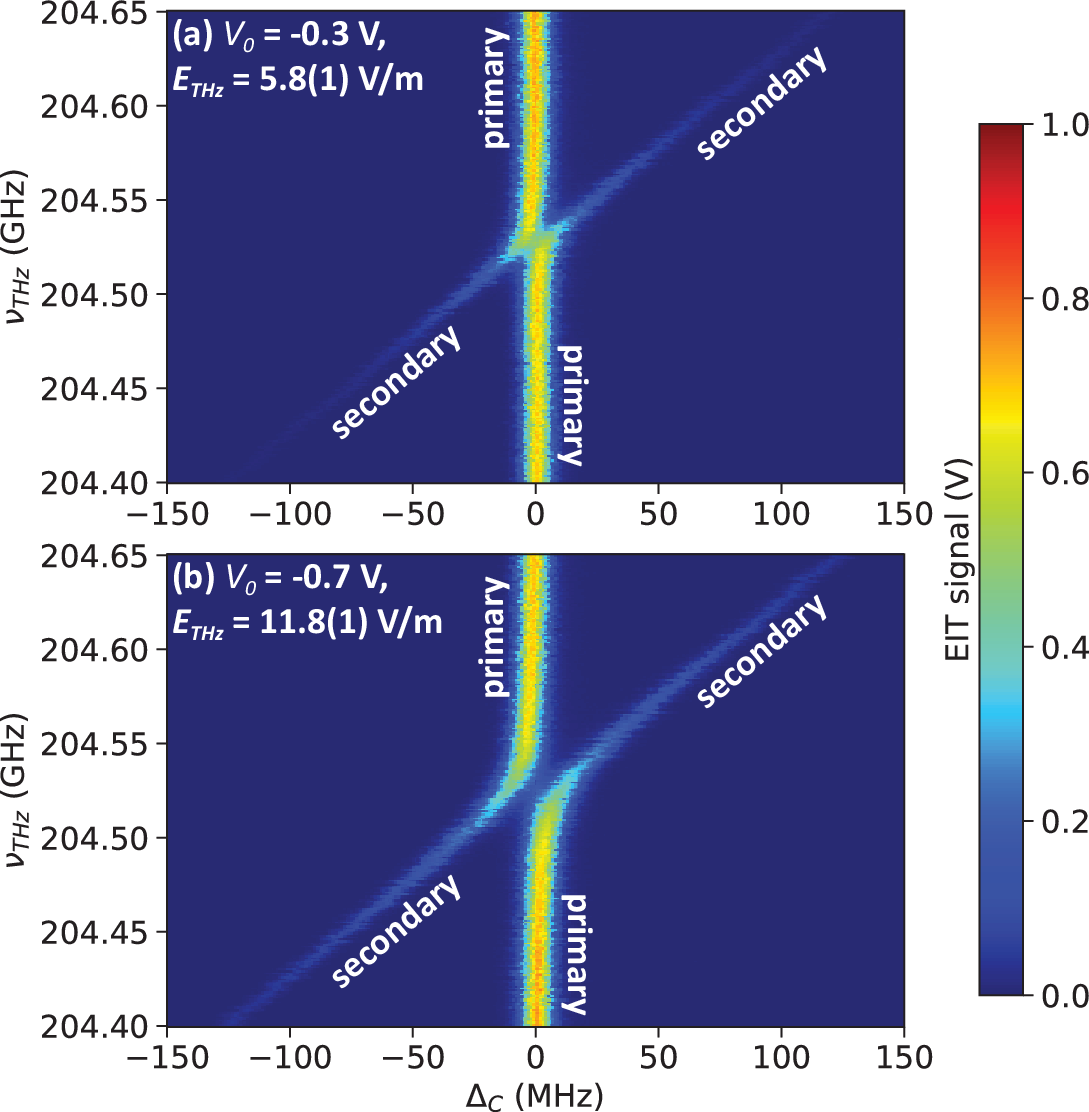}
\caption{Experimental EIT maps as a function of coupler laser detuning $\Delta_C$ and THz field frequency $\nu_{THz}$ for $V_0$ = -0.3~V~(a) and -0.7~V~(b). The THz field magnitudes, $E_{THz}$, extracted from the AT splittings at the $|17D_{5/2}\rangle\rightarrow|18P_{3/2}\rangle$ resonance using Eq.~\ref{eq:ETHz}, are indicated.}
\label{fig:map}
\end{figure}

We first calibrate the photomixer Tx output by measuring the electric field strength of the THz field, $E_{THz}$, versus $V_0$ in Eq.~\ref{eq:Vbias}, with $V_{mod} \equiv 0$, using the Rydberg-atom field sensor. 
The dc signal maps for $V_0=$~-0.3 and -0.7~V are shown in Fig.~\ref{fig:map}~(a) and~(b), respectively. In the presence of the THz field, the atomic spectrum exhibits two AT-split lines that are separated by the effective Rabi frequency $\Omega_{eff} = \sqrt{(\nu_{THz} -\nu_{THz,0})^2 + \Omega^2}$, with the THz Rabi frequency, $\Omega$, the resonant frequency, $\nu_{THz, 0}$, and the THz field frequency $\nu$. For $|\nu_{THz} -\nu_{THz,0}| \gg \Omega$, 
the stronger line,  denoted ``primary,'' is located approximately at the THz-field-free line position, while the weaker,  ``secondary'' line is located at $\Delta_C = \nu_{THz} -\nu_{THz,0}$. The secondary line has a slope of $d\nu_{THz}/d\Delta_C = 1$ when $|\Delta_C| \gg \Omega$, allowing us to calibrate the $\Delta_C$-axis based on wavemeter readings from the photomixer system. 

Approaching the resonance $\nu_{THz} = \nu_{THz,0}$, both lines in Fig.~\ref{fig:map} become more equal in area and undergo textbook-style AT-split avoided crossings. In the case of the higher THz field [Fig.~\ref{fig:map}~(b)], a weak background line at $\Delta_C = 0$ due the uncoupled $|17D_{5/2}, |m_J| = 5/2\rangle$ state is observed. Also, the AT-splitting near resonance exhibits a slight inhomogeneity due to the $\approx 20\%$ difference between the Rabi frequencies for  $|m_J| = 1/2$ and  $|m_J| = 3/2$, and potentially due to a minor inhomogeneity of the THz beam across the much smaller optical beams.

At resonance, the AT-split lines are equal in amplitude and area and are
are symmetrically split by an amount equal to the Rabi frequencies $\Omega_{m_J}$, 

\begin{equation}
\Omega_{m_J}= \frac{E_{THz} d_{m_J}}{h}
\label{eq:EThz_nobar} \quad,
\end{equation}

\noindent where $h$ is Planck's constant. The $m_J$-dependent EDTMs are $d_{mj} = d_{rad} \times w_{mj}$, with radial matrix element $d_{rad} = 312.0 e a_0$ ($e$ is the electron charge, and $a_0$ is the Bohr radius) and angular matrix elements $w_{1/2} = 0.490$ and $w_{3/2} = 0.400$~\cite{Reinhard2007}. 
Since the angular matrix elements are similar to each other, 
the AT-split lines for $|m_J|=1/2$ and $|m_J|=3/2$ are merged into a single AT-split line pairs in Fig.~\ref{fig:map} that correspond with the $m_J$-averaged EDTM, $\bar{d} = 139 e a_0$. In the average $\bar{d}$, the EDTM of the uncoupled states, $d_{5/2}=0$, is excluded. The separation between the symmetrically split AT peaks, determined using standard peak-fitting, equals the $m_J$-averaged on-resonant Rabi frequency for the THz-coupled states, $\bar{\Omega}$. Then the THz field, $E_{THz}$, follows from

\begin{equation}
E_{THz} = \frac{h \bar{\Omega}}{\bar{d}} \quad.
\label{eq:ETHz}
\end{equation}
\noindent 

Hence, measurement of the resonant splitting allows us to calibrate the THz electric field~\cite{Holloway2014}.

\subsubsection{Implementation}
\label{subsubsec:calidcimpl}

In detail, the atomic calibration data are acquired as follows. First, the THz field frequency, $\nu_{THz}$, is set by wavemeter-lock of the DFB laser frequencies. Its detuning from the $|17D_{5/2}\rangle\rightarrow|18P_{3/2}\rangle$ Rydberg resonance at $\nu_{THz,0}$ is $\Delta_{THz} = \nu_{THz} - \nu_{THz,0}$. The EIT spectra in the main cell and in the auxiliary EIT reference cell are simultaneously recorded on an oscilloscope while the coupler detuning $\Delta_C$ is scanned. The EIT peak in the THz-field-free auxiliary spectrum marks the location $\Delta_C=0$, allowing us to center all scans acquired with the main cell. Repetition of the procedure for a grid of $\nu_{THz}$-values, centered around the atomic resonance $\nu_{THz,0} = $~204.526(3)~GHz, and assembly of the data on the $(\Delta_C, \nu_{THz})$-plane yields full signal maps such as the ones shown in Fig.~\ref{fig:map}. The measurements are repeated for a set of fixed $V_{bias}=V_0$, with $V_{mod} \equiv 0$. The AT splittings observed at resonance, where the AT peaks are closest to each other, equal the THz Rabi frequency, $\Omega_{THz}$. Equation~\ref{eq:ETHz} then provides the THz electric field amplitude at the location of the field-sensing Rydberg atoms, $E_{THz}$.

As the THz field passes through the Pyrex vapor cell windows, it experiences THz absorption losses. These are quantified by recording the photomixer Rx photocurrent in its native mode, in which $V_{bias}$ is modulated, and the photocurrent is demodulated via lock-in detection. The photomixer Rx produces a photocurrent modulation amplitude $I_{THz, mod}$ that is proportional to the THz electric-field modulation amplitude (or the square root of the received THz power)~\cite{AD052026}. We measure the Rx photocurrent with and without the vapor cell in place. The ratio yields the THz field transmission coefficient, $\eta_{cell}$, through two cell windows. The THz field transmission for one cell window, $\eta_{cell,1} = \sqrt{\eta_{cell}}$. We found $\eta_{cell} = 0.77(2)$ and $\eta_{cell,1} = 0.88(1)$, in good agreement with previous reports~\cite{Rogalin2018}. The $E_{THz}$-values, obtained from the EIT spectra and Eq.~\ref{eq:ETHz}, are then corrected to obtain the THz field at the THz focus without the main cell in place, $E^{(0)}_{THz} = E_{THz}/\eta_{cell,1}$.

The dc calibration amounts to an analysis of the function $E^{(0)}_{THz}(V_0)$ with $V_{mod} \equiv 0$, plotted in Fig.~\ref{fig:calidc}. The function is approximately linear and is fitted with

\begin{equation}
E^{(0)}_{THz}  = k (V_{0}-V_{dcofs})\quad .
\label{eq:Ethz_cali}
\end{equation}

\noindent The fit parameter $k$ is the field calibration factor in units (V/m)/V. A dc offset $V_{dcofs}$ is included in the fit because $E^{(0)}_{THz}$ 
has a minute non-zero value when $V_0 = V_{mod} = 0$ in Eq.~\ref{eq:Vbias}. The offset $V_{dcofs}$ will be discussed further along with Table~\ref{tab1} in Sec.~\ref{subsubsec:calidcresults}.
Determination of $k$ and  $V_{dcofs}$ concludes the atomic calibration of 
the Terascan system in dc mode, where $V_{mod} \equiv 0$.

Furthermore, the power of the THz photomixer passing through the vapor-cell region (with the vapor cell removed), $P^{(0)}_{THz}$, can be obtained using the atom-based measurement of $E^{(0)}_{THz}$, such that

\begin{equation}
P^{(0)}_{THz} = \frac{\pi c \epsilon_0 w_{THz}^2}{4} [E^{(0)}_{THz}]^2 \quad.
\label{eq:P}
\end{equation}

\noindent There, $w_{THz}$ is the independently measured Gaussian beam waist parameter of the THz beam, $c$ is the speed of light, and $\epsilon_0$ is the permittivity of free space. 

\begin{table}[htb]
\caption{ Calibration results for continuous mode.}
\label{tab1}
\begin{tabular}{|c | c| c|} 
\hline
Quantity & Reference & Value \\
\hline
\makecell{$k$\\from non-forced fit} & \makecell{Eq.~\ref{eq:Ethz_cali} with $V_{bias,0}$\\as free parameter} & -17.6(9)~(V/m)/V \\
\hline
$k$ from forced fit & \makecell{Eq.~\ref{eq:Ethz_cali}\\with $V_{bias,0}=$~0} & -20.5(5)~(V/m)/V \\
\hline 
\end{tabular}
\end{table}

\subsubsection{DC calibration results}
\label{subsubsec:calidcresults}

\begin{figure}[htb]
\centering
\includegraphics[width=0.9\linewidth]{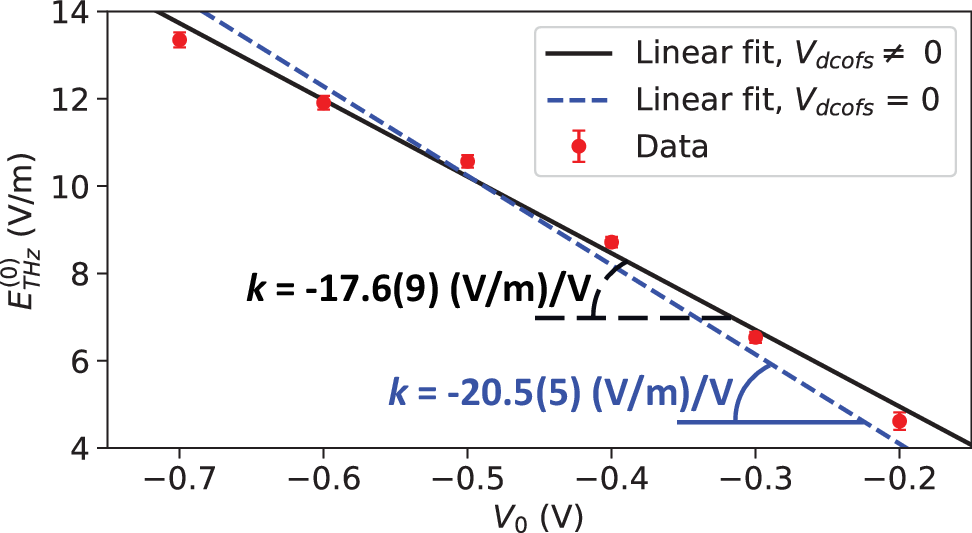}
\caption{Calibration of the THz transmitter-receiver system in dc mode. Data points are the THz electric field corrected for the transmission through the vapor cell wall, $E^{(0)}_{THz}$, vs bias voltage $V_0$ of the photomixer Tx. The linear fits shown (solid and dashed curves) and the indicated fitted slopes are based on Eq.~\ref{eq:Ethz_cali}.}
\label{fig:calidc}
\end{figure}

For the results shown in Fig.~\ref{fig:map}, we find $E_{THz} = 5.8(1)$~V/m in~(a) for $V_0=$~-0.3~V and $11.8(1)$~V/m in~(b) for $V_0=$~-0.7~V. When corrected for the field transmission through the vapor cell wall, $\eta_{cell,1}$, as described in Sec.~\ref{subsubsec:calidcimpl}, we obtain $E^{(0)}_{THz} = $~6.5(1)~V/m for $V_0=$~-0.3~V and $E^{(0)}_{THz} = $~13.4(2)~V/m for $V_0=$~-0.7~V, respectively. Results of $E^{(0)}_{THz}$ for a wider set of $V_0$-values are presented in Fig.~\ref{fig:calidc}.

Analysis along the lines of Sec.~\ref{subsubsec:calidcimpl} yields calibration data summarized in Table~\ref{tab1}. We perform a non-forced fit, in which both $k$ and $V_{bias,0}$ in Eq.~\ref{eq:Ethz_cali} are free fit parameters (solid line in Fig.~\ref{fig:calidc}), and a forced fit with $V_{bias,0} \equiv 0$ (dashed line). The  non-forced fit yields $k=-17.6(9)$~(V/m)/V and $V_{bias,0} = 0.08(2)$~V, while the forced fit yields $k=-20.5(5)$~(V/m)/V. The two calibration factors $k$ differ by 2 times their combined standard uncertainty range and are therefore reasonably close. The non-forced fit appears to be better. This finding indicates a slight non-linearity of the photomixer Tx, which may be caused by a variety of systematic effects~\cite{Lewis_2014, Peytavit2021, Ourednik:24}. 

Using Eq.~\ref{eq:P}, the calibrated output power of the photomixer Tx $P^{(0)}_{THz}$ is found to vary from 1.1(2) to 12(2)~$\mu$W over the range of $V_0$ used in our study. Although appearing on the lower side, considering potential losses along the THz beam line, this range is consistent with the specification of the utilized THz source~\cite{TopticaWP}. 

We note that the electric-field calibration at the vapor cell location depends on mirror reflectivities but is independent of focal lengths of the THz mirrors used if there are no significant effects from THz beam clipping and diffraction. The THz power, however, depends on both reflectivities and the focal length of the THz mirrors used.    

\subsection{Modulated mode}
\label{subsec:calimod}

\subsubsection{Methods and implementation}
\label{subsubsec:calimodmethod}

We next focus on calibrating the photomixer Rx in its native modulated mode.
We calibrate the Rx photocurrent magnitude without the Pyrex vapor cell in place, $I^{(0)}_{THz}$, versus $V_{mod}$. This calibration step does not involve atoms and depends on the alignment of the THz system. According to the manufacturer's data, 
\begin{equation}
I^{(0)}_{THz} = \sigma V_{mod}\quad,
\label{eq:Ithz_calimod}
\end{equation}
\noindent where $\sigma$ is the calibration factor in units of nA/V.

\begin{figure}[t!]
\centering
\includegraphics[width=0.9\linewidth]{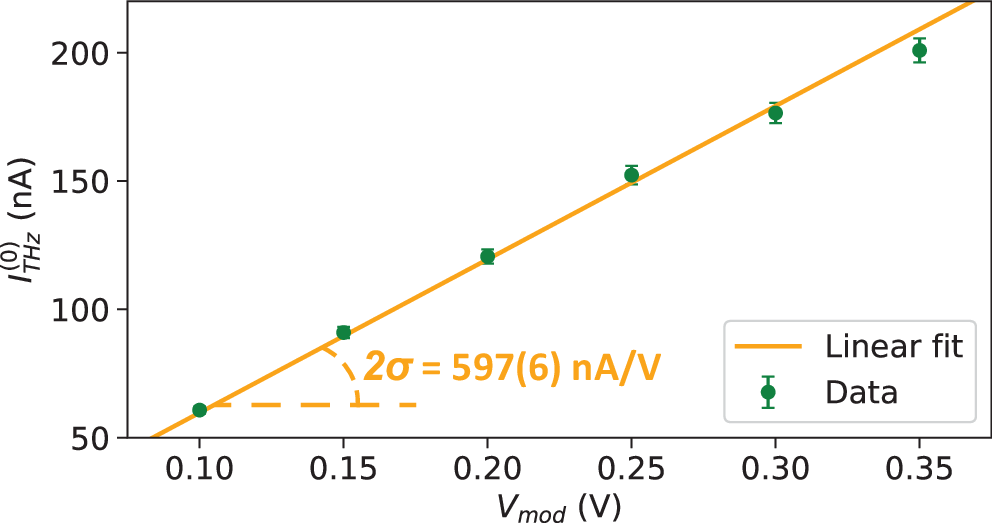}
\caption{Calibration of the THz transmitter-receiver system in its native modulation mode. Data points are the THz-induced current reading of the photomixer receiver, $I^{(0)}_{THz}$, as a function of the transmitter modulation amplitude, $V_{mod}$, for fixed $V_0 = $~-0.4~V and with the vapor cell removed. Solid curve is a linear-fit result using Eq.~\ref{eq:Ithz_calimod}. The obtained fitted slope is indicated.}
\label{fig:calimod}
\end{figure}

The results are shown in Fig.~\ref{fig:calimod}, together with the linear fit using Eq.~\ref{eq:Ithz_calimod}. For the experimental data, we have determined the Terascan's background current as a function of $V_{mod}$ by blocking the THz beam and recording the Rx reading. While the background current slightly increases with  $V_{mod}$, it always stays well below 1\% of the signal current $I^{(0)}_{THz}$. Therefore, it has no significant effect in Fig.~\ref{fig:calimod}. The error bars shown reflect the variation of  $I^{(0)}_{THz}$ within the frequency range of interest (the vertical range in Fig.~\ref{fig:map}). 
We finally determine the electric-field responsivity $\gamma$ of the photomixer receiver Rx in units of nA/(V/m) using Eqs.~\ref{eq:Ithz_calimod} and~\ref{eq:Ethz_cali},
\begin{equation}
\gamma = \frac{I^{(0)}}{E^{(0)}} =  \frac{\sigma}{|k|}\quad.
\label{eq:EoverI}
\end{equation}

\begin{table}[htb]
\caption{ Calibration results for modulated mode.}
\label{tab2}
\begin{tabular}{|c|c|c|} 
\hline
Quantity & Reference & Value \\
\hline
$\sigma$ & Eq.~\ref{eq:Ithz_calimod} & 597(6)~nA/V \\
\hline
\makecell{$\gamma$ using $k$\\from non-forced fit} & Eq.~\ref{eq:EoverI} & 34(2)~nA/(V/m) \\
\hline
\makecell{$\gamma$ using $k$\\from forced fit} & Eq.~\ref{eq:EoverI} & 29(1)~nA/(V/m) \\
\hline
\end{tabular}
\end{table}

\subsubsection{Modulated-mode calibration results}
\label{subsubsec:calimodresults}

Following Sec.~\ref{subsubsec:calimodmethod}, the calibration factor in Eq.~\ref{eq:Ithz_calimod} is found to be $\sigma = 597(6)$~nA/V. For $\gamma$, we provide results for both the non-forced and the forced fits used to determine $k$. Modulated-mode calibration data are summarized in Table.~\ref{tab2}.

We note that the calibration factor $\sigma$ depends on the overall quality of the Terascan system alignment. The calibration factor $\gamma$, which relies on atomic calibration, also depends on the focal length of the THz mirrors used.   

\section{Discussion}
\label{sec:disc}

We conclude the work with a discussion of several types of relevant noise sources.
The Terascan system's photomixer Rx noise could, in principle, be important. Assuming the photocurrent transimpedance amplifier has an impedance of $R = 50$~$\Omega$, the classical Johnson noise is given by $I_{J} = \sqrt{4 k_B T B / R}\approx 33$~pA, where $k_B$ is the Boltzmann constant, $T$ is temperature, and $B$ is the detector bandwidth (3.3~Hz throughout this study). This translates into $E_{min} \approx 1.1$~mV/m of THz field noise when using our atomic calibration factor 29(1)~nA/(V/m). In comparison, the measured current noise is approximately 48~pA, corresponding to $E_{min} \approx$~1.7~mV/m. The difference of a factor of approximately 1.5 indicates the presence of a small amount of technical noise, such as interferometric phase noise in the photomixer Rx. The photomixer Rx noise corresponds with electric fields that are orders of magnitude smaller than the coherent THz electric fields. 

Expressed as a field-noise spectral density, the Johnson-current-noise limit corresponds to $I_J/\sqrt{B}=\sqrt{4 k_B T/R}\approx 18$~pA/$\sqrt{\rm Hz}$, or $E_J\approx 0.6$~mV/m/$\sqrt{\rm Hz}$ using the same (forced) atomic calibration factor. The measured Rx current noise corresponds to approximately $0.9$~mV/m/$\sqrt{\rm Hz}$. Thus, although the photomixer Rx characterized here operates within a factor of about 1.5 of its classical Johnson-noise limit, its field-equivalent noise floor remains substantially above demonstrated Rydberg atomic receiver sensitivities in the mm-wave band that can reach below $10~\mu$V/m/$\sqrt{\rm Hz}$~\cite{Legaie.2024}. At these performance limits, our benchmarking demonstrates that the atom-based measurement provides not only an SI-traceable calibration of the incident 204-GHz field, but also an absolute-field benchmark for classical receiver sensitivity performance, highlighting the potential advantage of Rydberg receivers for weak-field detection and calibrated field measurements in the mm-wave and THz regimes.

Next, we consider noise due to black-body background radiation (BBR). The BBR noise power is given by the black-body spectral irradiance integrated over the bandwidth and the area of the sensor. Conventional pyroelectric and bolometric THz power sensors typically have apertures on the order of 1~cm$^2$ and bandwidths in the THz range. This leads to a 300-K BBR power $\gtrsim 1 \mu$W that must be subtracted from the reading, resulting in low accuracy at low THz powers. Our unattenuated THz photomixer Tx yielded a power reading of $\sim 26$~$\mu$W on such a sensor, which is larger than the highest obtained $P^{(0)}_{THz}$ in Sec.~\ref{subsubsec:calidcresults} by a factor of approximately 2. We did not observe changes in the bolometric power measurements as $V_0$ was varied, indicating that these were obstructed by the BBR background. Also, the bolometric measurements were susceptible to changes in environmental conditions and variability of nearby heat sources. In contrast, the photomixer Rx used generates an output proportional to the THz electric field (not power)~\cite{AD052026}. It has a sub-mm$^2$ aperture and a bandwidth of only about 3.3~Hz (given by the utilized single-acquisition sampling rate). At 204~GHz, its 300-K BBR power is $\lesssim 10^{-19}$~W, which is orders of magnitude below the relevant Tx powers. The Rydberg atomic THz sensor is also subject to BBR electric-field background, which is $\sim 5~\mu$V/m/$\sqrt{{\rm{Hz}}}$, or about 5~mV/m, when integrated over the $\sim 1$-MHz sensitivity bandwidth of the atomic sensor. This is orders of magnitude below our calibration fields. Our estimates of BBR effects show that both photomixer Rx and atomic THz sensors were operated well above their respective BBR noise floors. This conclusion elevates the value of our atomic calibration procedure.

\section{Conclusion}
\label{sec:concl}
We have demonstrated SI-traceable calibration of a commercial THz photomixer Tx and Rx system using Rydberg atomic electrometry. The atomic measurement allows direct determination of the amplitude of the THz electric field emitted by the transmitter as a function of its bias and modulation voltages. The calibration of the photomixer Rx is afforded by simultaneous detection of photomixer Rx current and atomically measured electric fields. The presented approach provides a practical method for benchmarking Tx and Rx systems in the mm-wave and THz regimes, including characterization of systematic effects from atmospheric attenuation~\cite{SLOCUM201349} and optical properties of THz components that may be present in the systems~\cite{Dai:04, braakman2011, alligood2013, harrison2018}. Our work lays out pathways towards absolute calibration of THz measurement systems and future sensitivity comparisons between atomic and classical THz detectors. 

\section*{Acknowledgments}
\label{sec:acknowledgments}
We would like to thank Dr. Luís Felipe Gonçalves, Dr. James E. Carey, Dr. Teng Zhang and Dr. Bineet Dash for useful discussions. This work was supported by Rydberg Technologies Inc. and, in part, by the Defense Advanced Research Projects Agency (DARPA) EQSTRA Program under agreement No. HR00112530134. Distribution Statement “A” (Approved for Public Release, Distribution Unlimited). The views and conclusions contained in this document are those of the authors and should not be interpreted as representing the official policies, either expressed or implied, of the U.S. Government.

\bibliographystyle{apsrev4-2}
\bibliography{bibliography.bib}

\end{document}